\documentclass[twocolumn]{aastex63}

\received{}
\revised{}
\accepted{}
\submitjournal{ApJ}

\shorttitle{Lessons from SN Refsdal}
\shortauthors{Zitrin A.}

\begin{document}

\title{Lessons from the first multiply imaged supernova: A revised Light-Traces-Mass strong lensing model for the galaxy cluster MACS J1149.5+2223}

\correspondingauthor{Adi Zitrin}
\email{adizitrin@gmail.com}

\author[0000-0002-0350-4488]{Adi Zitrin}
\affiliation{Physics Department,
Ben-Gurion University of the Negev, P.O. Box 653,
Be'er-Sheva 84105, Israel}



\begin{abstract}
Our light-traces-mass (LTM) strong-lensing model for MACS J1149.5+2223 has played several key roles over the last decade: it aided the identification of multiple images in this cluster and the study of MACS1149-JD1 at redshift $z\simeq9$, it was used to estimate the properties of the first multiply imaged supernova, Refsdal, in its discovery paper, and of the first caustic crossing event by a cluster, Lensed Star 1. Supernova Refsdal supplied an invaluable opportunity to conduct a blind test of the ability of common lens-modeling techniques to accurately describe the properties of SN Refsdal's images and predict the reappearance of one of its counter images that was due about a year post-discovery of the original Einstein cross. Thanks to this practice, in which our submitted model yielded some outlying results, we located a numerical artifact in the time delay (TD) calculation part of the code, which was now fixed. This artifact did not influence the reproduction of multiple images (i.e., the deflection fields -- which are those constrained directly from the observations) or the derived mass model, and so it remained unnoticed prior to supernova Refsdal, emphasizing the importance of blind tests in astronomy. Here we update our model and present revised LTM measurements for Refsdal. These are important not only for completing the LTM view of the Refsdal event, but also because they affect the range of values predicted from different lens-modeling techniques and thus the range of systematic uncertainties for the TD calculation and the resulting Hubble constant. 

\end{abstract}



\section{Introduction}\label{sec:intro}
Measuring the expansion rate of the Universe and the Hubble constant in particular has been one of modern astronomy's Holy Grails. 

More than 50 years ago, \citet{Refsdal1964MNRAS} suggested a way to measure the Hubble constant using a multiply imaged supernova (SN). Since light rays to each multiple image traverse a different path and cross a different potential well, each image has a different, delayed arrival time. The difference in arrival times, i.e., the time delay difference (which we refer to here, in short, simply as the time delay; TD), is inversely proportional to the Hubble constant. Hence, by measuring the TD between multiple images of the exploding SN, a measurement of the Hubble constant would be enabled (assuming a lens model is at hand, e.g., from the position of the observed SN images). In fact, Refsdal's idea applies to a wider range of observable transients. More commonly, thanks to their brightness and the fact that they are not short lived, multiply imaged quasars -- typically lensed by galaxies and monitored over a long time-span -- have been used to derive accurate estimates of the Hubble constant \citep[e.g.,][the latter finding  $H_{0}=73.3^{+1.7}_{-1.8}$ km/s/Mpc]{Vuissoz200CosmograilTDH0,Suyu2013TDs,Wong2019H0}. Such measurements have recently gained an additional interest as, together with constraints from improved Cepheid measurements to Type Ia SNe \citep[e.g.,][$H_{0}=74.03\pm 1.42$,  km/s/Mpc]{Riess2019}, they seem to be in tension with the values derived from Planck's observations of the cosmic microwave background \citep[CMB;][]{Planck2018Params}, which favor a smaller value of $67.4\pm0.5$ km/s/Mpc. This discrepancy could possibly hint to new physics \citep[e.g.,][]{Blandford2020}, but various systematics, such as those manifested in the discrepancy between the local value from Cepheids and that from Tip of the Red Giant Branch calibration \citep[][$H_{0}=69.8 \pm0.8\pm 1.7$ km/s/Mpc]{Freedman2019H0}, have to be understood first.

Due to their relative short duration and typical rate, up until a few years ago multiply imaged SNe have not, in fact, been detected at all and thus were not used to constrain $H_{0}$. About five years ago and fifty years after Refsdal's original prediction, the first clear example of a multiply imaged SN was discovered \citep{Kelly2015Sci}, as an Einstein cross around a cluster member in the field of the galaxy cluster MACS J1149.5+2223 (M1149 hereafter, \citealt{EbelingMacs12_2007}). As a courtesy to the original paper, the SN was dubbed \textit{SN Refsdal}. A couple of other multiply imaged SNe have also been detected: an apparently bright SN, multiply-imaged by a galaxy was noted in the year preceding Refsdal \citep[PS1-10afx;][]{Quimby2013BrightSN,Quimby2014SciLensedSN}, albeit it was not resolved into multiple images and a TD was not measured for it; and the first multiply imaged Type Ia SN, also lensed by a galaxy, was later discovered \citep[iPTF16geu;][]{Goobar2017SciLensedTypeIa}. 
SN Refsdal appears to have exploded in an arm of a spiral galaxy multiply imaged by the cluster, M1149, and further, more locally, by a cluster member. Being clearly resolved and lensed by a cluster so that an additional appearance of the SN was predicted roughly a year after the detection of the Einstein cross \citep{Kelly2015Sci,Kelly2016reappearance}, SN Refsdal opened a unique door for both predicting when a future SN explosion could be observed, and, for measuring $H_{0}$. In addition, it allowed for a true blind comparison test of various lens modeling techniques  -- or the parametrization of the underlying mass density -- to accurately predict the reappearance \citep{Treu2016Refsdal,Rodney2016RefsdalComp,Kelly2016reappearance}.

The galaxy cluster M1149 was detected thanks to its X-ray signal as part of the MAssive Cluster Survey (MACS; \citealt{EbelingMacs12_2007}). The first lensing analysis of M1149 was published around the same time by \citet{ZitrinBroadhurst2009}, who proposed many of the gold multiple images in this cluster (see Table \ref{multTable}), and by \citet[][]{Smith2009M1149}, who also measured spectroscopically the redshift for some of these systems. Following its compelling features, M1149 was also observed in 16 bands as a high-magnification cluster in the Cluster Lensing And Supernova survey with Hubble treasury program \citep[CLASH;][]{PostmanCLASHoverview}, and included in the subsequent ultra-deep Hubble Frontier Fields program \citep[HFF;][]{Lotz2017HFF}, and in the Grism Lens-Amplified Survey from Space \citep[GLASS;][]{Treu2015GLASS}. It has also been a part of extensive ALMA programs, such as the ALMA Frontier Fields survey (PI: Bauer; e.g., \citealt{Gonzalez2017}), or the ALMA Lensing Cluster Survey (ALCS; PI: Kohno). Following this rich coverage, including ground based spectroscopy (e.g., \citealt{Treu2016Refsdal} and references therein) many other lens models for the cluster have been published to date \citep [e.g.,][see also the HFF webpage \footnote{\url{https://archive.stsci.edu/prepds/frontier/lensmodels/}}]{Rau2014M1149,Richard2014FF,Johnson2014HFFmodels,Kawamata2016modelsHFF,Diego2016refsdal}.

M1149 has been the at the heart of a wide variety of studies, from high-redshift galaxies \citep[e.g.,][]{Zheng2017HFF1149,Ishigaki2018HFF} or ultra diffuse galaxies \citep[e.g.,][]{Janssens2019}, through lensed galaxy properties \citep[e.g.,][]{Wang2017metal1149,MunozArancabia2018,Carvajal2020} and their ionizing photon budget \citep[e.g.,][]{Emami2020ionizing1149}, to finding AGN in cluster fields \citep[e.g.,][]{DellaCosta2020AGN1149}; as a few random examples. In addition, M1149 is known to host several record-breaking phenomena: aside from lensing the first resolved multiply imaged SN, it has been found, for example, to host one of the highest redshift galaxies known \citep{Zheng2012Nature} later verified spectroscopically to become the farthest Lyman-alpha emitter to date \citep{Hashimoto2018}; it allowed the plausible detection of a $\sim10^{10}$ M$_{\odot}$ black hole near the center of the brightest cluster galaxy \citep[BCG;][]{Chen2018BHm1149}; and it allowed the first detection of a cosmological caustic crossing event by a cluster, dubbed MACS J1149 Lensed Star 1 \citep[MACS1149-LS1; also known as Icarus;][]{Kelly2017CC}. The latter was detected in follow-up observations of SN Refsdal, and consists, most likely, of a young massive star in an arm of the same spiral galaxy hosting Refsdal at $z=1.49$ that crosses the caustics \citep{Kelly2017CC}. This detection has opened the door for finding more such transients in other clusters that were relatively frequently visited with HST \citep{Chen2019MACS0416CCE,Rodney2018Trans0416,Kaurov2019MACS0416CCE}. 

Over the past decade our Light-Traces-Mass (LTM) model for M1149, updated various times throughout, has played some key roles in studies of this cluster, from finding some of the first multiple images \citep{ZitrinBroadhurst2009}, through aiding in qualifying it for the HFF via simulations of high-redshift galaxy expectations \citep[section 3 in][]{Lotz2017HFF}, to helping in the investigation of MACS1149-JD1 \citep{Zheng2012Nature}, SN Refsdal \citep{Kelly2015Sci}, or MACS1149-LS1 \citep{Kelly2017CC,Diego2017CC}. Our goal here is to present a long-due update of the model. In particular, we take advantage of the blind comparison work by \citet{Treu2016Refsdal,Rodney2016RefsdalComp,Kelly2016reappearance} that examined the ability of common lens modeling techniques to predict the reappearance and properties of SN Refsdal, and in which our submitted model yielded some outlying estimates -- especially regarding the Einstein cross. This experiment has led us to detect a very minor, but apparently critical, numerical inconsistency in the TD calculation part in our code. We fix this so-called error and show that the model we submitted blindly before does in fact reproduce a reasonable TD surface for SN Refsdal. We update the model, and publish here new LTM measurements for SN Refsdal. While preliminary estimates of the TD have been presented in \citet{Treu2016Refsdal,Rodney2016RefsdalComp,Kelly2016reappearance}, no such information is used in the minimization. Finally, we briefly discuss the implications on the Hubble constant, $H_{0}$, of our new TD measurement  \citep[e.g.,][]{Vega-Ferrero2018RefsdalH0,Grillo2018H0m1149}.

M1149 will also be included in various JWST/GTO programs\footnote{\url{https://www.stsci.edu/jwst/observing-programs/approved-gto-programs}}, and more exciting science is thus expected in this cluster in the near future. Our model is made publicly available\footnote{\url{https://www.dropbox.com/sh/mac3vuian8gyjkq/AADTD0ENfIwclkjp9GJHnUhua?dl=0}}. A future version of the model, over a larger field of view, is expected in the framework of the Beyond Ultra-deep Frontier Fields and Legacy Observations \citep[BUFFALO;][]{Steinhardt2020BUFFALO} program with Hubble.


The paper is organized as follows: in \S \ref{s:code} we give a short overview of the LTM methodology and in \S \ref{s:modeling} we outline our modeling of M1149. In \S \ref{s:results} we present and discuss the results from this modeling with an emphasis on SN Refsdal, namely the magnification ratios, TDs, and implications for the Hubble constant. The work is concluded in \S \ref{s:summary}. Throughout this work we use a $\Lambda$CDM cosmology with $\Omega_{M}=0.3$, $\Omega_{M}=0.7$, and $H_{0}=70$ km/s/Mpc. Unless otherwise stated, errors are $1\sigma$, and we generally use AB magnitudes \citep{Oke1983ABandStandards}.

\section{The LTM code}\label{s:code}
The LTM code relies on the assumption that light traces mass, so that the weighted luminosity distribution of cluster galaxies can act as a guide for the shape of the total matter distribution. A first version of the code was written and used by \citet{Broadhurst2005a} to find dozens of multiple images in Abell 1689, revealing a very large lens. Based on their success, \citet{Zitrin2009_cl0024} developed a new, simpler LTM code that included a minimal number of free parameters. This code was since constantly improved, and used to identify multiple images and model dozens of galaxy clusters. Perhaps the greatest value of the code lies in that it is frequently successful in predicting the location of multiple images based only on the luminosity distribution of cluster members, even before it is constrained with any multiple image systems \citep[][see also \citet{Zalesky2020AstroLens}]{Carrasco2020}. A second importance is that it is distinct from typical, analytic lens modeling techniques and thus probes a different range of solutions. We give here a brief overview of the code; for more details see \citet[][see also \citealt{Carrasco2020}]{Zitrin2014CLASH25}.

The starting point of the model is the distribution of cluster members and their luminosities. Each galaxy is assigned with a power-law mass density profile, scaled in proportion to its luminosity. The power-law exponent is a free parameter of the model and the same for all galaxies. A grid is defined -- typically matching a portion of the HST image of the cluster -- and the total deflection field and mass density distribution of the galaxies are calculated on that grid. This galaxies' map is then smoothed (in Fourier space) with a Gaussian kernel to produce a map of the dark matter component. The width of the Gaussian is a free parameter as well. The deflection field maps from the galaxies and the DM are then added with a relative weight, which is also a free parameter of the model, and supplemented by a two component external shear whose strength and position-angle parameters are typically left free to be optimized as well. The overall noramlization of the model is the sixth free parameter.

In addition, we often leave free the core size, relative weight (i.e., the relative mass-to-light ratio), ellipticites and position angles of key cluster members, such as the BCGs. All other member galaxies are assumed to be circular and core free, for simplicity.

The optimization of the model is carried out by minimizing a $\chi^2$ function that measures the distance of predicted multiple images from their observed location, via a Monte-Carlo Markov Chain with a Metropolis-Hastings algorithm. Punishing terms can be added for images with wrong parity or if extra images are predicted. We also include some annealing in the procedure, and the chain typically runs for several thousand steps after burn-in phase. Errors are typically calculated from the same Markov chain.

\begin{figure*}
 \begin{center}
  \includegraphics[width=0.8\textwidth,trim=0cm 0cm 0cm 0cm,clip]{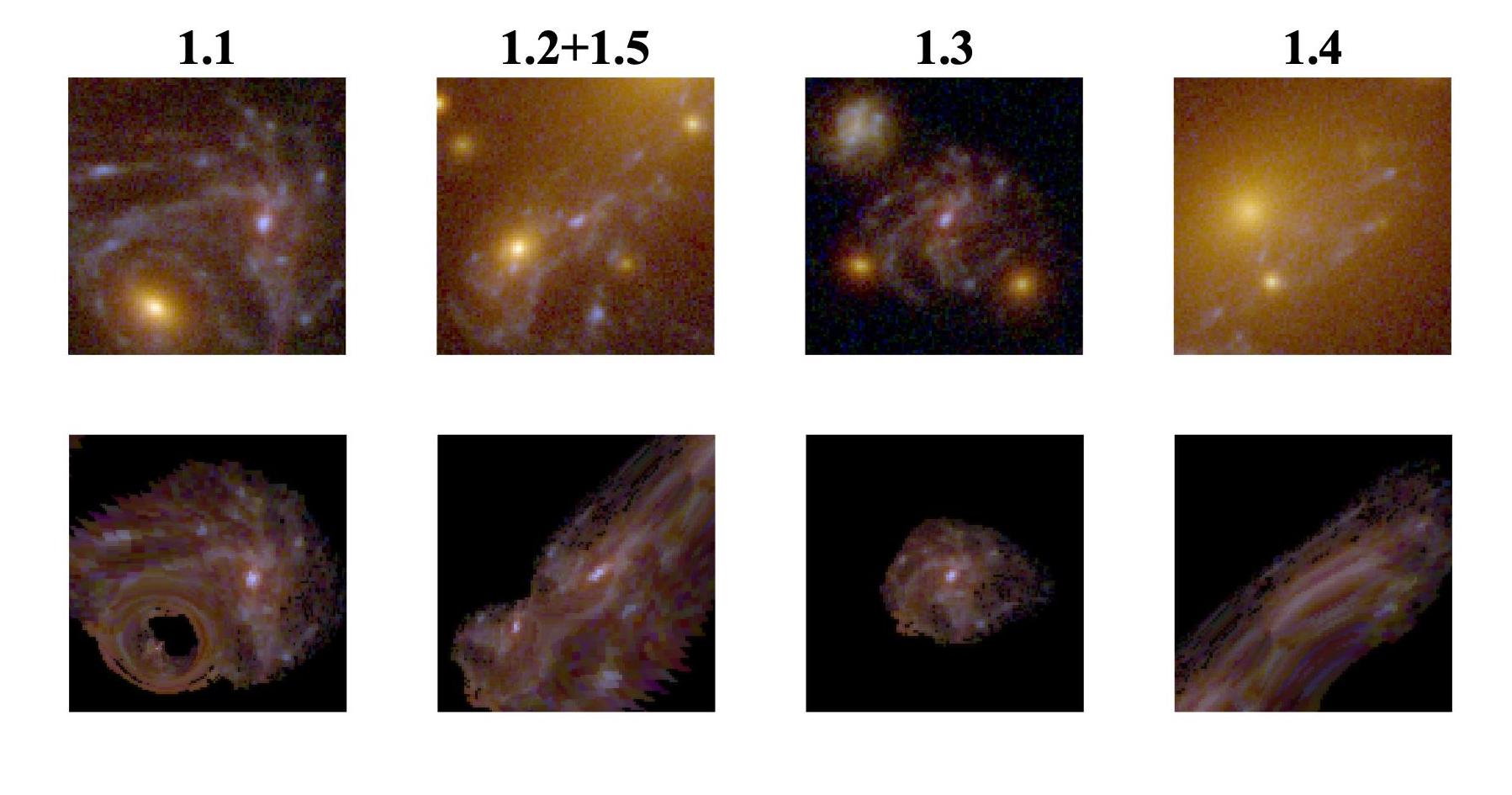}
 \end{center}
\caption{Reproduction of system 1 by our model. We send image 1.1 to the source-plane and back to obtain the reproduction of the other images of this system. The first stamp is 7.8\arcsec on 7.8\arcsec and the three other stamps are 6.5\arcsec on 6.5\arcsec.}\label{fig:Sys1Spiral}
\end{figure*}

\section{LTM modeling of M1149}\label{s:modeling}
As mentioned, the first SL LTM model for M1149 was constructed in 2009 with an old code, prior to CLASH multiband data, and before redshifts were available for this cluster. This model was later updated with improvements to the code and using CLASH data or newer products agreed on within the HFF framework, and included strong lensing, or strong$+$weak lensing constraints combined \citep[e.g.][]{Zitrin2014CLASH25,Treu2016Refsdal,Zheng2017HFF1149}. Here, we update the model submitted to the \citet{Treu2016Refsdal}, \citet{Kelly2016reappearance} and \citet{Rodney2016RefsdalComp} blind tests, with a corrected TD calculation (see \S \ref{ss:older}). 

Since the model is limited to the grid resolution and our goal here is to examine the properties of SN Refsdal, we concentrate on the central $\sim140\arcsec \times140\arcsec$ area of the cluster. The final model has the same spatial resolution as the model presented in \citet{Treu2016Refsdal}, of 0."065/pix, native to the public CLASH HST images (although some parts in the minimization are done in a few-times lower resolution for speed-up purposes). 

We leave the weight of four galaxies free: the central BCG (RA=11:49:35.70, DEC=+22:23:54.71); a second bright cluster member at RA=11:49:37.55, DEC=  +22:23:22.49; another bright member next to the center at RA=11:49:36.86, DEC=+22:23:46.97; and the galaxy that lenses SN Refsdal into the Einstein cross (RA=11:49:35.47, DEC=+22:23:43.63). The ellipticity and position angle of the main BCG are left free to be optimized around their values measured by SExtractor \citep{BertinArnouts1996Sextractor}, and the ellipticity and position angle of the second and third BCGs are kept fixed on their measured values. A core is assumed for the central BGC, whose size is a free parameter. The general galaxy power-law parameter, $q$, is fixed to $1.45$, based on various trials that have shown that this value works very well in reproducing the partial spiral counter image north of the BCG (see Fig. \ref{fig:Sys1Spiral}).

The multiple image constraints we use here are given in Table \ref{multTable}. These consist mainly of the \emph{gold} multiply imaged systems presented in \citet{Treu2016Refsdal} and \citet{Finney2018M1149}, with some \emph{silver} counter images that our model predicts as secure. The \emph{gold/silver/etc.} ranking was given based on voting by lens modelling teams in the framework HFF and SN Refsdal's efforts \citep[][for details]{Treu2016Refsdal}. We also adopt the detailed mapping of the main, lensed spiral galaxy's internal knots as presented in \citet[][and references therein]{Treu2016Refsdal} and \citet{Finney2018M1149}, and use these as additional constraints, as listed in Table \ref{knotTable}. All systems were fixed to their spectroscopic redshift where available. System 6 \& 7 do not have a spectroscopic measurement and their redshift was fixed to $\simeq2.6$ based on their photo-$z$. We do not use any measured TDs or magnification ratios as input.

For the $\chi^2$ calculation, a positional uncertainty of 0.5'' is adopted for most multiply images, but for the four SN images in the Einstein cross we adopt 0.1''. We also include punishing terms to guarantee the correct parity of these images, and to verify that no additional (full) image of the spiral galaxy is predicted just north of the BCG, which was the case in some trial models.

\begin{figure*}
 \begin{center}
  \includegraphics[width=0.49\textwidth,trim=0cm 0cm 0cm 0cm,clip]{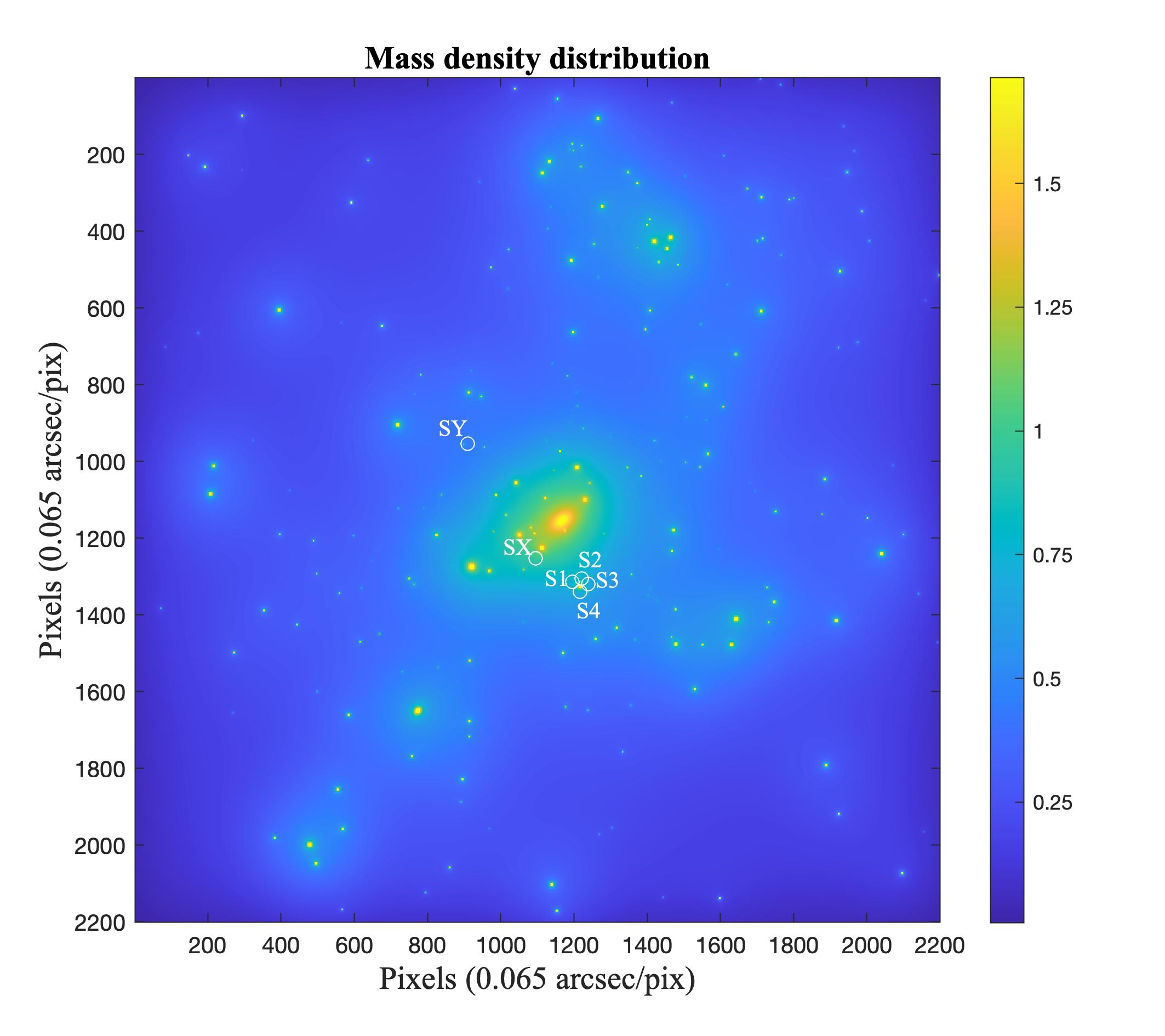}
    \includegraphics[width=0.49\textwidth,trim=0cm 0cm 0cm 0cm,clip]{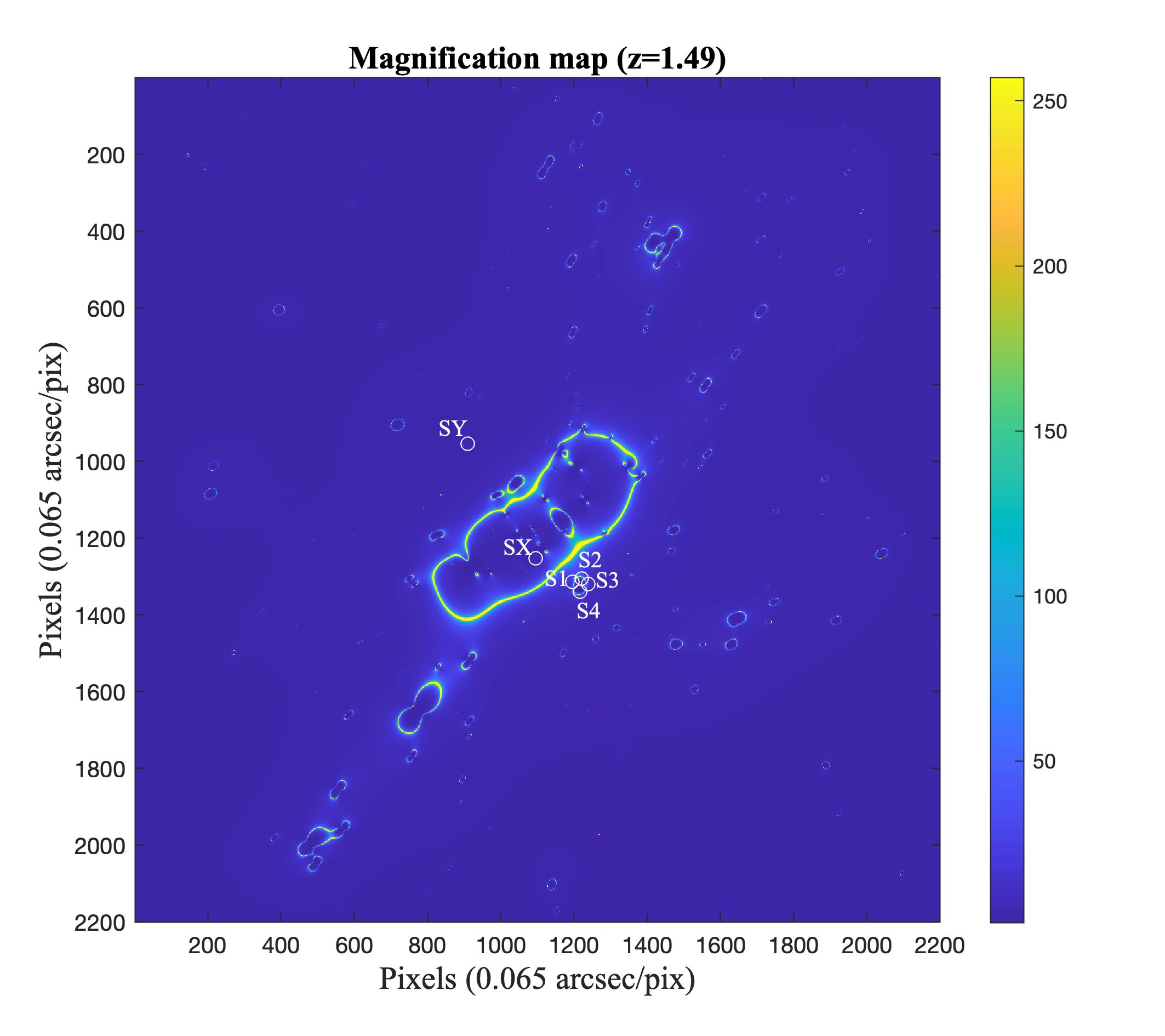}\\
      \includegraphics[width=0.49\textwidth,trim=0cm 0cm 0cm 0cm,clip]{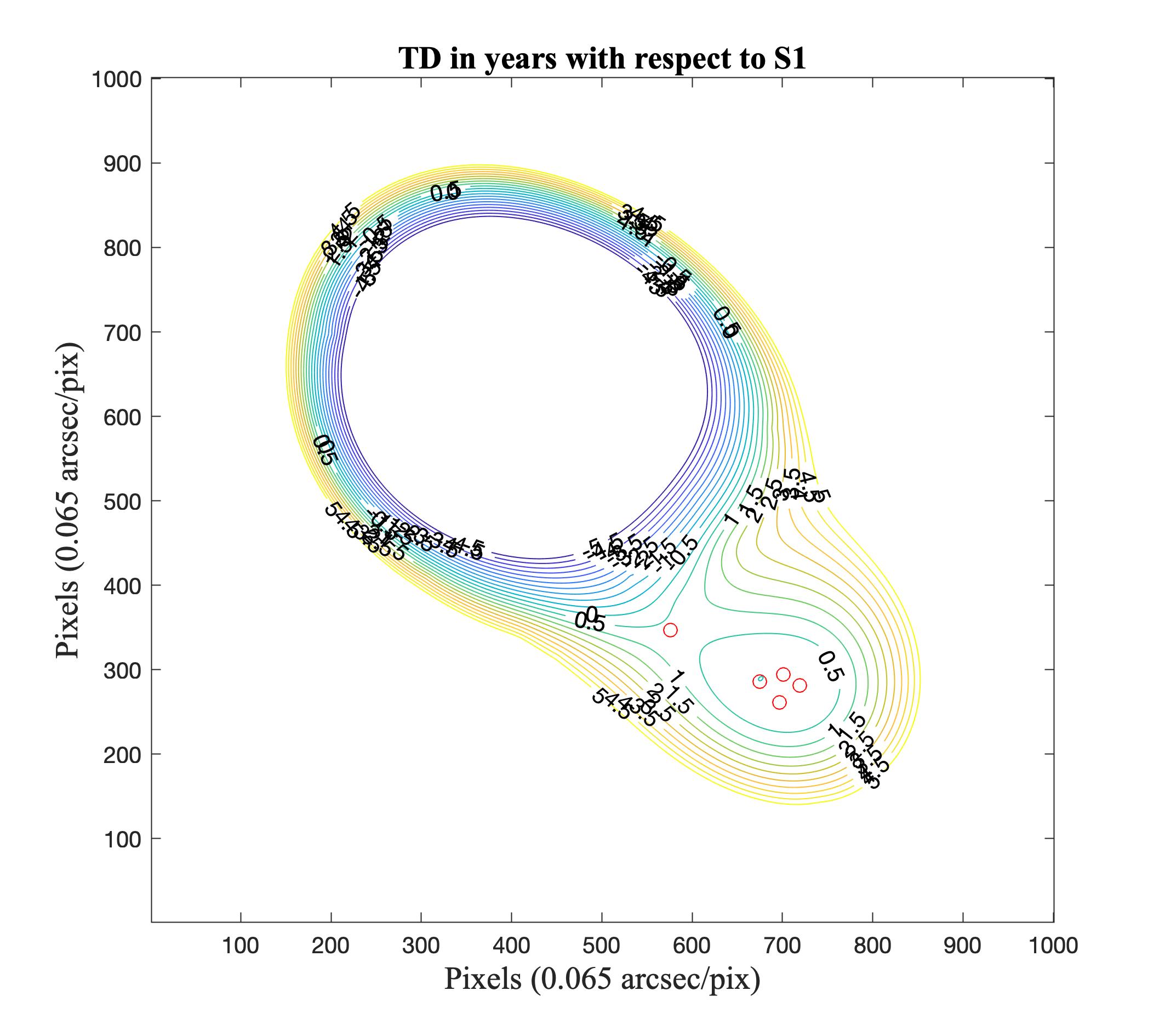}
    \includegraphics[width=0.49\textwidth,trim=0cm 0cm 0cm 0cm,clip]{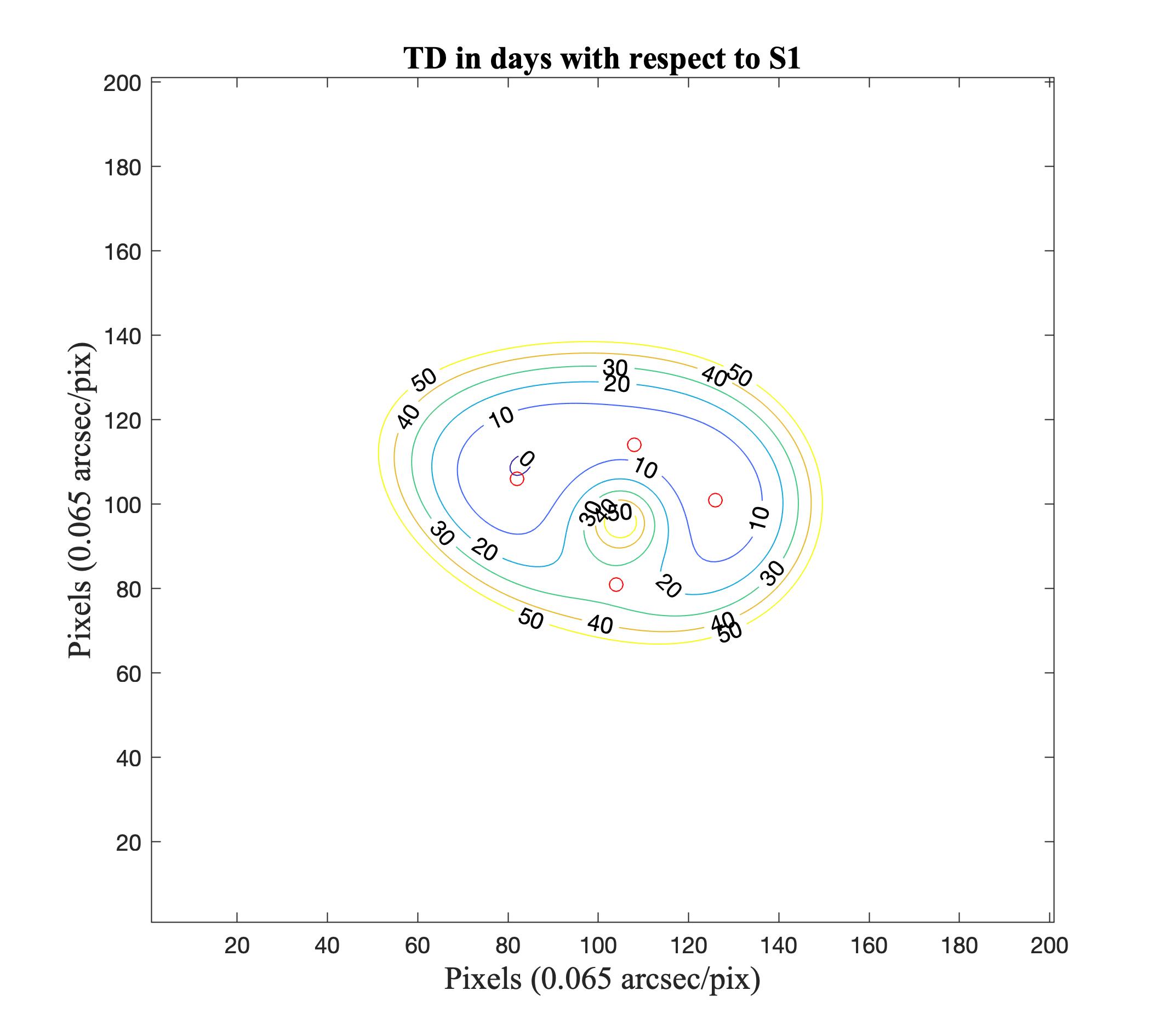}
 \end{center}
\caption{Our revised LTM SL model for M1149. \emph{Upper left} figure shows the surface mass density, $\kappa$, for the redshift of system 1, the spiral galaxy at $z=1.49$, with the positions of S1-S4, SX, and SY, marked with white circles; the \emph{Upper right} figure shows the magnification map for that redshift, similarly marking the SN image positions;  the \emph{Bottom left} figure shows contours of the TD surface with respect to the SN image S1, in decrements of 0.5 years, marking with red circles the Einstein cross images and and SX;  the \emph{Bottom right} figure shows contours of the TD surface with respect to the SN image S1, in decrements of 10 days, marking with red circles the positions of the Einstein cross images.}\label{fig:kappa_mag_TD}
\end{figure*}

\section{Results}\label{s:results}

Our resulting mass model is seen in Fig. \ref{fig:kappa_mag_TD}, along with the magnification map and TD surface contours. The final model has an image reproduction \emph{rms} of $0.68\arcsec$, and a $\chi^2\simeq242$. With the many multiple images used, the number of constraints is $N_{c}=170$, and the number of free parameters $N_{p}=11$, such that the number of degrees of freedom is DOF$=N_{c}-N_{p}=159$. Hence the \emph{reduced} $\chi^2$ is $\simeq1.5$. We note that for the spiral knots alone the \emph{rms} is a bit lower, $0.49\arcsec$ (note also the \emph{rms} is limited by the grid's resolution; a quantification of this effect will be done elsewhere). 

The reproduction of the complex configuration of system 1 can also constitute a (more qualitative, in our case) testimony for the model's credibility in the central region. We show the reproduction of this system in Fig. \ref{fig:Sys1Spiral}, compared to the data, by sending the largest spiral image to the source-plane and back. The reproduction of some knots is not very prominent, perhaps indicative of some inaccuracy or a resolution limit, although overall the model seem to predict very well the detailed appearance of the images. 

\subsection{LTM TDs and magnifications for SN Refsdal}

The model's TDs and magnification ratios, with respect to S1, for SN Refsdal's five observed images (S1-S4 and SX), as well as the older unobserved SN image (SY), are given in Table \ref{TDs}. The quoted errors were derived using a 100 random models from a designated MC chain, and we list both the 68.3\% and 95\% confidence intervals. To account for (some) systematics, these were extracted using an effective $\sigma_{pos}\simeq1.7\arcsec$, which we have found better captures the range of values from different trial models constructed in the course of modelling M1149. The original statistical intervals are about 3-5 times smaller. The relevant values seem to broadly agree, in most cases, with the early measurements by \citet{Rodney2016Refsdal} and \citet{Kelly2016reappearance}, listed in the same table as well. However, we anticipate that updated measurements for SN Refsdal, exploiting a wider monitoring time-span of Refsdal's appearances, will become available in the future. While a major discrepancy between such updated measurements and our current estimate may persist, we note that our numbers seem to very broadly agree with the range of estimates from the different models seen in \citet{Rodney2016Refsdal}, \citet{Treu2016Refsdal}, and \citet{Kelly2016reappearance}. Most models predict TDs of order days for S2 and S3 with respect to S1; about 15-30 days between S4 and S1, and from $\sim220$ to $\sim380$ days between SX and S1. Nevertheless, it seems that LTM TD estimates, and most notably for the SX-S1 TD, are systematically lower than those obtained by most parametric techniques. In addition, if updated SX-S1 TD measurements remain significantly different than our estimate, it can potentially cast important clues on properties of the underlying matter distribution that are critical for TD mapping. Based on the early measurements by \citet{Kelly2016reappearance} the TD is not very likely to be shorter than our estimate (see also \citealt{Baklanov2020}), and is more likely to be significantly larger: following \citet{Kelly2016reappearance} SX was detected on 11 December 2015, and shows fainter traces also on 14 November 2015  -- almost exactly a year after detecting the Einstein cross on 10 November 2014. In contrast, observations from 30 October 2015 yielded no statistically significant detection of SX. Hence, as seen in Fig. 3 in \citet{Kelly2016reappearance}, unless the magnification of SX is very small compared to S1, most chances are that the SX-S1 TD is larger than our estimate of $\sim$267 days (or $\sim224$ days, depending on the source position).

\begin{figure*}
 \begin{center}
  \includegraphics[width=0.99\textwidth,trim=6cm 0cm 3cm 0cm,clip]{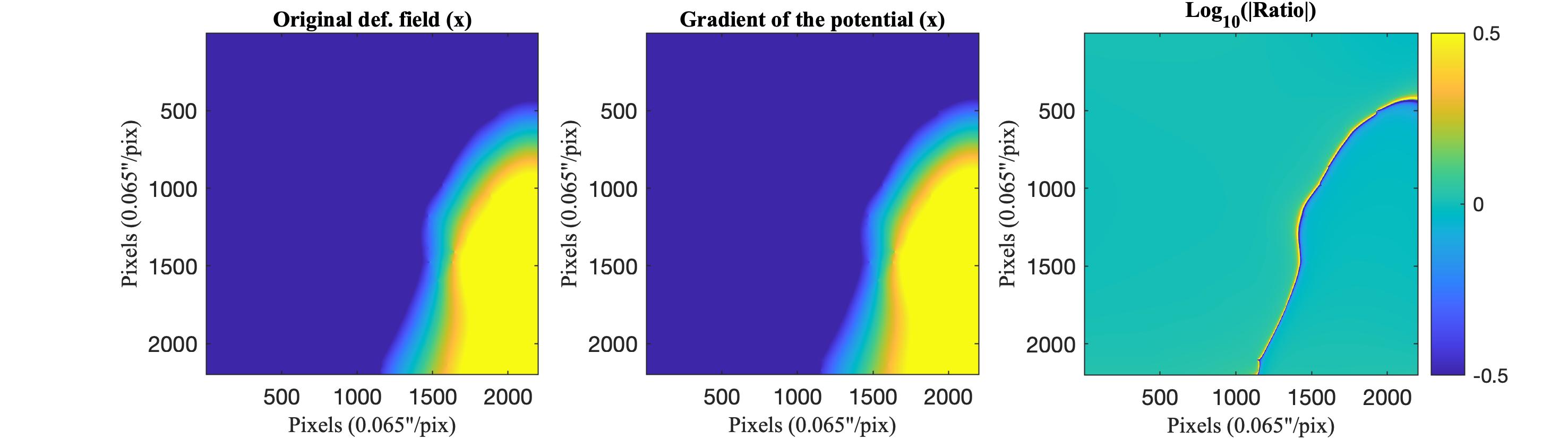}
 \end{center}
\caption{Residual from the calculation of the numerically-integrated potential. \emph{Left:} The original best-fit x-axis deflection field. \emph{Center:} The derived x-axis deflection field obtained from the gradient of the (numerically-integrated) potential. \emph{Right:} $\log_{10}$ of the absolute value of the ratio of the two deflection fields. While the ratio equals unity in most of the field, a numerical residual is clearly seen. Values were limited to $\pm$0.5 to emphasize the artifact.}\label{fig:artifact}
\end{figure*}

\begin{figure*}
 \begin{center}
  \includegraphics[width=0.49\textwidth,trim=0cm 0cm 0cm 0cm,clip]{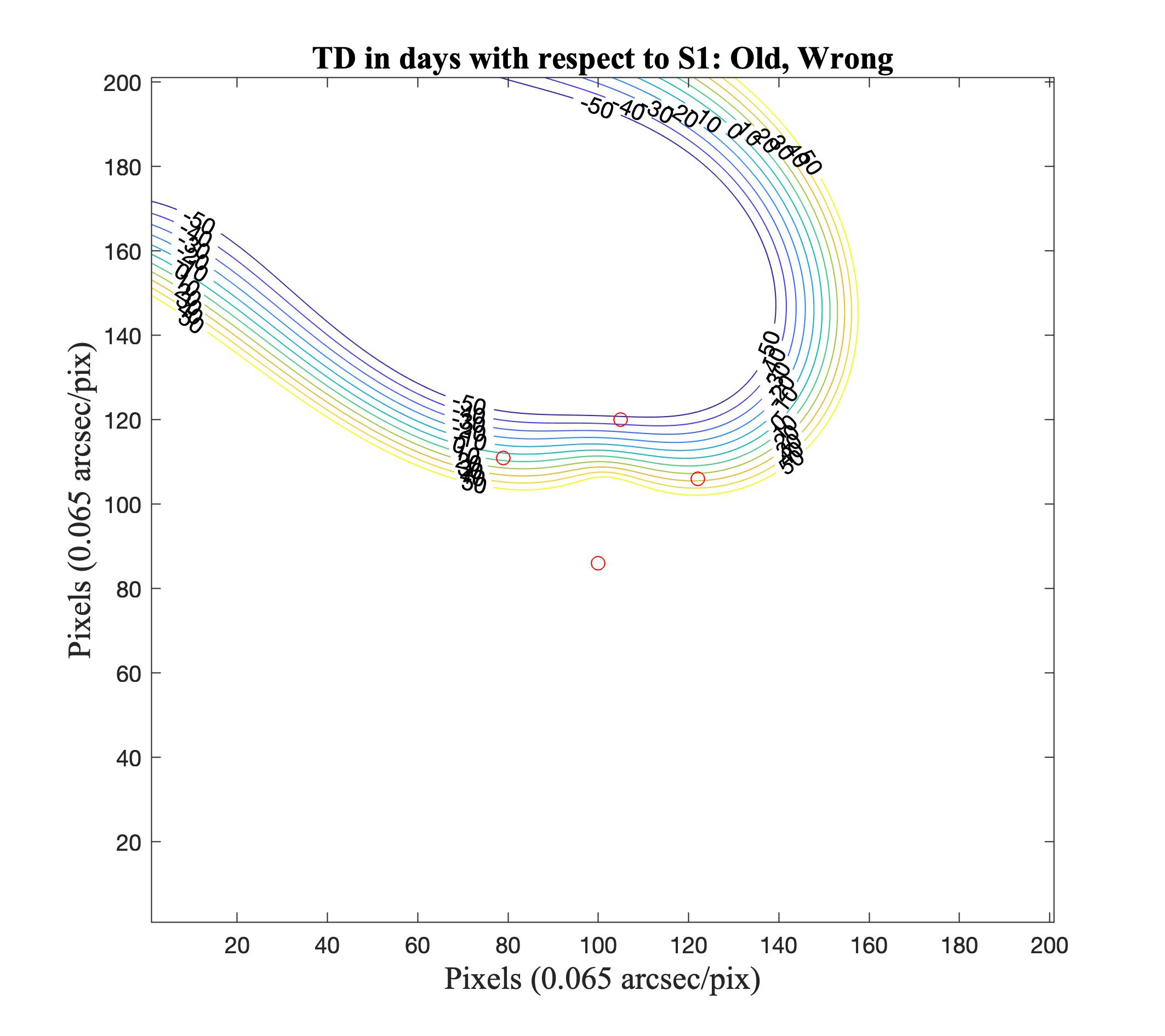}
    \includegraphics[width=0.49\textwidth,trim=0cm 0cm 0cm 0cm,clip]{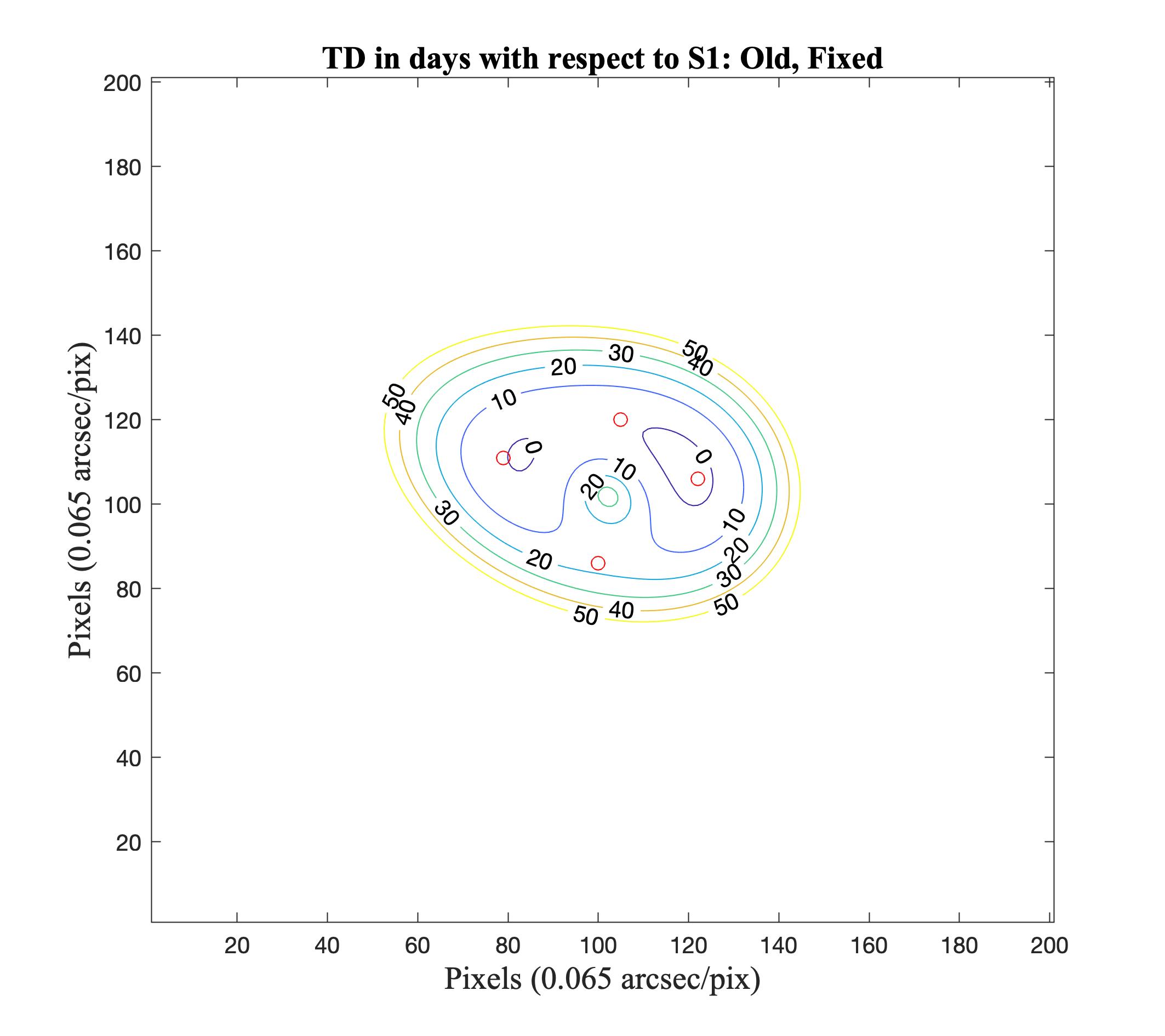}
 \end{center}
\caption{Contours for the TD surface from our old LTM model, in decrements of 10 days. \emph{Left}: Example of the deformed TD surface with the old code in which the potential and deflection fields were not perfectly self-consistent (potential obtained by numerical integration). \emph{Right}: TD surface for the same old model, after correction to the code (the deflection field was taken directly to be the gradient of the (integrated) potential, see text). Red circles mark the positions of the four SN images of the Einstein cross. The TD surface from our new model presented in this work is shown in Fig. \ref{fig:kappa_mag_TD}. and the relevant values are listed in Table \ref{TDs}.}\label{fig:WrongFixed}
\end{figure*}

The absolute magnifications are not listed in Table \ref{TDs}, and instead given here. Our model predicts magnifications of $\mu\sim20-22$ for the images S1-S3, $\mu\simeq5.2$ for S4, and $\mu\simeq4.7$ for SX. SX is predicted at RA=177.39995, DEC=22.39664 by our model, about $\simeq0.4\arcsec$ from its observed location in \citet{Kelly2016reappearance}. The model suggests that SY appeared $\simeq18$ years ago, with a magnification of $\mu\sim3.4$.

It is worth mentioning that in the presented model, the effective ellipticity of the galaxy that forms the Einstein cross was not optimized separately; the galaxy was modeled as a simple circular power-law, and the ellipticity of its critical curve, which produces the four images of the SN, was automatically obtained as such likely due to the shear caused by the main potential well, centered on the BCG. It is conceivable that an independent modeling of this lensing galaxy could yield more refined results. Similarly, we remind the reader that the LTM code is limited to the grid's resolution, which could affect the TD measurement. To assess the typical magnitude of the effect, we repeat the TD measurements from our model with a resolution twice lower in each axis. The TDs between the images of the Einstein cross change by about few to $\sim10$ days (and so the estimate for the order of images S1-S3 can also change), and the TD between SX and S1, as well as the confidence interval, increase by about a month. As for the magnification ratios, while most typically agree within 1-2$\sigma$ with the higher resolution result, some may be more severely affected -- especially those images near the critical curve (i.e., the Einstein cross images) and the higher-resolution results are needed to obtain more stable predictions for them. In cases where very high-resolution results are needed, analytic codes (which essentially can obtain any desired resolution) may be preferable, although in some cases higher-resolution grids could also be employed with LTM.

Note that since the TDs are affected by the exact position of the source (e.g., \citealt{BirrerTreu2019}), we quote in Table \ref{TDs} values for two cases: in the first the source position is taken as the average delensed position of the four SN images of the Einstein cross, similar to what was done prior to the appearance of SX. For comparison, and although SX is not used in the minimization, we also list the estimates if the source position is taken as the mean delensed position of both the four Einstein cross images and SX. As can be seen, most properties agree fairly well (although sometimes marginally) within the errors between the two scenarios, but for some cases the effect on the TD and arrival order can be more significant. The TDs quoted by the first source position is probably more representative of the properties of S1-S4, whereas it remains to be seen if the TD taken from the second source position including SX, describes better the SX-S1 TD, as one may speculate. In that sense the \emph{first} SX-S1 TD [68.3\%][95\%], 224.4 [221.4 -- 272.7] [197.8 -- 305.5] days, is the one that would have been given prior to the detection of SX, had we used the current model -- and it seems that it matches almost exactly the old LTM TD estimate in \citet{Kelly2016reappearance}, but with a lower SX/S1 magnification ratio.

Finally, we briefly mention that, by design, the range of different trial models we constructed in the process of re-modeling M1149 (specifically models with rms$<$1\arcsec), spanning a range of input configurations (number of freely weighted galaxies and which, ellipticity of the BCGs, number of systems with free redshifts, etc.), typically fall within the range of errors for our best-fit model presented in Table \ref{TDs}. The majority of these models give values for the SX-S1 TD of $\sim170$ days to $\sim340$ days, most concentrated around 200-260 days, but few -- usually the less well constrained models -- give larger or smaller TDs. 

\begin{deluxetable*}{lllll}
\tablecaption{Time Delays and Magnification ratios for SN Refsdal}
\label{TDs}
\tablecolumns{5}
\tablewidth{0.9\linewidth}
\tablehead{
\colhead{Parameter
} &
\colhead{$\Delta t(t)$
} &
\colhead{$\Delta t(p)$
} &
\colhead{LTM$^{c}$ [68.3\% CI] [95\% CI]
} &
\colhead{LTM$^{d}$ [68.3\% CI] [95\% CI]
}
}
\startdata
$\Delta t_{S2:S1}$ & $4 \pm 4^{a}$ & $7 \pm 2^{a}$ & 5.4 [3.3 -- 6.4] [2.8 -- 7.2]& 2.3 [0.8 -- 5.2] [-0.2 -- 7.1] \\
$\Delta t_{S3:S1}$& $2 \pm 5^{a} $&$0.6 \pm 3^{a}$& 1.6 [0.8 -- 2.1] [0.6 -- 2.9] & -10.0 [-11.5 -- -2.8] [-12.9 -- -0.43]\\
$\Delta t_{S4:S1}$& $24 \pm 7^{a}$& $27 \pm 8^{a}$& 26.3 [23.4 -- 27.4] [22.6 -- 28.3]& 12.9 [10.8 -- 19.5] [7.75 -- 22.2]\\
$\Delta t_{SX:S1}$& $345 \pm 10^{b}$& $345 \pm 10^{b}$ & 224.4 [221.4 -- 272.7] [197.8 -- 305.5] & 267.4 [263.1 -- 305.3] [249.8 -- 320.3]\\
$\Delta t_{SY:S1}$& ---&  --- & -6522 [-6623 -- -6137] [-6834 -- -6025]& -6335 [-6479 -- -6083] [-6776 -- -5768]\\
$\mu_{S2}/\mu_{S1}$& $1.15 \pm 0.05^{a}$ & $1.17 \pm 0.02^{a}$& 0.86  [0.69 -- 1.34] [0.45 -- 1.57]& (0.86  [0.69 -- 1.34] [0.45 -- 1.57])\\
$\mu_{S3}/\mu_{S1}$& $ 1.01 \pm 0.04^{a}$ & $1.00 \pm 0.01^{a} $& 0.94 [0.88 -- 1.00] [0.78 -- 1.09]& (0.94 [0.88 -- 1.00] [0.78 -- 1.09])\\
$\mu_{S4}/\mu_{S1}$& $ 0.34 \pm 0.02^{a} $ & $0.38 \pm 0.02^{a}$& 0.23 [0.19 -- 0.32] [0.14 -- 0.36]& (0.23 [0.19 -- 0.32] [0.14 -- 0.36])\\
$\mu_{SX}/\mu_{S1}$& $0.28\pm0.1^{b}$& $0.28\pm0.1^{b}$ &0.21 [0.19 --  0.23] [0.16 -- 0.25] & (0.21 [0.19 --  0.23] [0.16 -- 0.25])\\
$\mu_{SY}/\mu_{S1}$ & --- & --- & 0.15 [0.13 -- 0.17] [0.11 -- 0.18]& (0.15 [0.13 -- 0.17] [0.11 -- 0.18])\\
\enddata
\tablecomments{Early measurements of SN Refsdal's TDs (in days) and magnification ratios, along with estimates from our new LTM model. Note that TDs and magnifications were not used as constraints in the minimization, and similarly, any information regarding SX (and SY) was not used explicitly as well. We expect more accurate measurements of SN Refsdal's properties will become available in the future. \\
$^{a}$ - Taken from Table 3 in \citet{Rodney2016Refsdal} based on a set of templates (t), or on polynomials (p) [this notation is adopted from \cite{Grillo2018H0m1149}].\\ 
$^{b}$ - Taken from \cite{Grillo2018H0m1149}, adopted therein following Fig. 3 in \citet{Kelly2016reappearance}. In practice, the 1$\sigma$ uncertainties in \citet{Kelly2016reappearance} are closer to 10\%.\\ 
$^{c}$ - Source position taken as mean of the delensed positions of the Einstein cross images, S1-S4. These values are those relevant for the properties of the four Einstein cross images.\\
$^{d}$ - Source position taken as mean of delensed position of five images: the Einstein cross images, S1-S4 and the position of SX.\\
The magnification values are independent of the exact source position.
}
\end{deluxetable*}

\subsection{Comparison with the older model}\label{ss:older}

Differences with respect to our older LTM models that appeared in \citet{Treu2016Refsdal}, \citet{Rodney2016Refsdal}, and \citet{Kelly2016reappearance}, can be divided into three main categories: differences in the input and the constraints that were used, minor corrections to the code, and the TD calculation fix.

One difference in input between the older model and the current one, although perhaps not very significant, is that we rerun the photometry of cluster members to create a new member galaxy catalog. Some minor differences exist also for the constraints: in the previous model all multiple image families in the gold list were used except system 14, and a partial list of knots of the spiral galaxy were used, whereas here we use the full gold list (with few silver images in gold systems) and including the full list of knots from \citet[][see also \citet{Treu2016Refsdal}]{Finney2018M1149}. Systems with no spectroscopic redshift were previously left to be freely optimized in the old model, whereas here these systems (namely systems 6, and 7) were kept fixed on their typical best-fit photometric redshift, $z\simeq2.6$. Image position uncertainties were adopted to be $0.5\arcsec$ for most systems also in the older model, and for the four SN images for which $0.15\arcsec$ was used previously, we now used $0.1\arcsec$. 

One technical difference between the older model and the current one is that the efficiency of the code was slightly improved, so it could be actually run in the native HST resolution whereas before it was typically run with a few times lower resolution and then interpolated to the native pixel scale (or with higher resolution but less MC steps). 

Several other minor corrections to the code have taken place since our 2014 model, but there was no significant change to the modeling scheme.  The main improvement in the current model estimates, however, is the fix in the TD calculation part. Thanks to the blind comparison undertaken for SN Refsdal \citep{Treu2016Refsdal,Rodney2016Refsdal,Kelly2016reappearance}, we detected a problem in the previously submitted TD surface: the SN images did not seem to form at the extrema of the Fermat surface as expected (see Fig. 8 in \citet{Treu2016Refsdal}, Fig. \ref{fig:WrongFixed} left here). After investigating the origin of the problem, we reached the following conclusion -- and corresponding fix. In the LTM formalism many of the operations are performed in Fourier space for speed up purposes. For example, the deflection fields are calculated in Fourier space, and since in our scheme the potential cannot be calculated analytically or parametrically from the mass map (or deflection field), it is instead derived by integrating, in Fourier space, the mass density distribution times the log of the distance to each point (see eq. 51 in \citealt{NarayanBartelmann1996Lectures}). The potential due to the external shear is then added analytically onto the same grid to obtain the total potential. The total potential was then used together with the deflection fields to estimate the TD surface. The problem was that the gradient of this potential resulted in deflection field components (or if differentiated again, kappa maps) that were verified to be overall very similar to the original deflection fields or kappa map that were integrated to begin with, but in fact were not \emph{exactly} the same (see Fig. \ref{fig:artifact}). Dividing the original deflection fields and kappa map by their parallels obtained via the potential, we get that the mean of the division maps is $\simeq1$ to within 1\% in all cases, but that about 5-10\% of the pixels in each division map are outliers, differing by more than 5\% or 10\%, respectively. This discrepancy caused the TD surface to deform (Fig. \ref{fig:WrongFixed} left; the effect is similar, essentially, to shifting the source position). To fix this we simply override the "original" deflection field in each step with the deflection field obtained by taking the gradient of the potential. Hence, the potential in each step now constitutes the new starting point, from which all components of the model are then self-consistently deduced. With this the TD surface, \emph{including of our older model submitted to the original comparison test}, forms where expected with the SN images in its extrema.

It should also be mentioned that in retrospect, the error could have been in principle found also without SN Refsdal, just by examining the TD surface on small scales or with sufficiently detailed self-consistency tests. In practice, however, it was the blind comparison and more importantly the fact that our older results for the Einstein cross (i.e., on small scales especially; \citealt{Rodney2016Refsdal}) were outliers, that made us notice this issue, eventually.

\subsection{Implications for the Hubble constant}

The TD equation is given by (e.g., eq. 63 in \citealt{NarayanBartelmann1996Lectures}):
\begin{equation}\label{eq:TD}
    t(\vec\theta)=\frac{1+z_{l}}{c}\frac{D_{l}D_{s}}{D_{ls}}\left[\frac{1}{2}(\vec\theta-\vec\beta)^2-\psi({\vec\theta})\right]
\end{equation}
where $\vec\theta$ is the image position, $\beta$ is the source position as indicated by the lens model, and $\psi$ the gravitational potential given by the lens model. $D_{l}$, $D_{s}$, and $D_{ls}$ are the angular diameter distances to the lens, to the source, and between the lens and the source, respectively. The above equation describes the delayed arrival time of each image compared to an undeflected light ray from the same source had there not been any intervening lens, and we denoted the arrival time difference between the different multiple images simply as the TD. The combination of angular diameter distances, $\frac{D_{l}D_{s}}{D_{ls}}$ -- sometimes called the TD distance, especially if including the ($1+z_{l}$) factor as well  --  is inversely proportional to the Hubble constant. Hence, using eq. \ref{eq:TD}, by comparing observed TDs with the TDs expected from a lens model that was constructed using the positions of the multiple images (note the model is independent of the Hubble constant; the latter only enters when estimating the TD), an estimate of the true underlying Hubble constant can be derived. The TD distance depends much more mildly also on the cosmological parameters \citep{Grillo2020TDaccuracy1149}, and a full statistical treatment is needed for proper constraints on the Hubble constant, to include the explicit statistical uncertainties from both the model and TD measurement. Nevertheless, one can obtain a rough estimate by simply rescaling the model's TD using the measured TD: 
\begin{equation}\label{eq:TD2}
H_{0,true}=\text{TD}_{model}/\text{TD}_{true} \cdot H_{0,model} 
\end{equation}
where we used in our modeling $H_{0,model}=70$ km/s/Mpc. The final, ''true'' measured TD for Refsdal is yet unknown at the time of writing. If, for example, we use the TD$_{true}=345\pm10$ between S1 and SX adopted by \citet{Grillo2018H0m1149}, the Hubble constant implied by our model would be as low as $\sim$46 km/s/Mpc, with a 95\% CI of $\simeq$[40 -- 62] km/s/Mpc, in the case where the cross' source position is adopted; or $\sim$54 km/s/Mpc with a 95\% CI of $\simeq$[51 -- 65] km/s/Mpc, in the case where the mean cross+SX source position is adopted. However if the final measured TD would be around 220-270 days (this can be the case if the magnification of SX were sufficiently low), then the Hubble constant would be around 70 km/s/Mpc (for the two source positions, respectively), as is measured from other probes and other TD systems as well. For previous measurements of the Hubble constant with SN Refsdal we refer the reader to \citet{Vega-Ferrero2018RefsdalH0} and \citet{Grillo2018H0m1149}, where discussions of possible systematics and the accuracy of the TDs was investigated, for example, by \citet{Williams2019TDaccuracy1149}, \citet{BirrerTreu2019} and \citet{Grillo2020TDaccuracy1149}.

\section{Conclusions}\label{s:summary}

The main purpose of this paper was to present a long-due revision of our LTM model for M1149, following an inconsistency (a numerical artifact) that was discovered in the TD calculation part of the code. The artifact was found after noticing that the TD surface of the LTM model we submitted to the SN Refsdal blind challenge in 2015 was defomred, not producing the multiple images at its extrema. We have found that the origin of the problem was a slight inconsistency between the deflection field or kappa maps, and the integrated potential, which led to, effectively, a slightly shifted source position and deformed TD surface. Note that the inconsistency involves only the TD estimates through the calculation of the potential, and as it does not affect the multiple image reproduction or the accuracy of the LTM model more generally, it remained unnoticed (and mostly irrelevant) prior to SN Refsdal. We revise our code to use the potential as starting point in each step so that all lensing quantities are self-consistently derived from it directly. We rerun our model for M1149, presented here, and update the TDs and magnifications for SN Refsdal accordingly. These, in turn, could be compared to updated measurements of SN Refsdal. 

If a large discrepancy remains between our current TDs and those measured for SN Refsdal (for example, while they very broadly agree, the LTM predictions for SX-S1 TD in particular, seem to be systematically lower than those from most parametric techniques), it will enable us to test which specific assumptions in our model lead to a distinct TD, and may cast important clues regarding which properties of the underlying matter distribution are crucial for correct TD mapping. In parallel to this paper we also publish a \emph{parametric} model for M1149, to enable a more direct comparison (Zitrin, A., submitted and posted online).

Our work highlights the extended usefulness of blind prediction tests in astronomy.

\section*{acknowledgements}
This work was made possible by useful discussions with Pat Kelly, Tommaso Treu, and Steve Rodney. I am very grateful for these discussions and for their comments on the manuscript. I am also very grateful for Tommaso Treu's notice of the deformed TD surface, which led us to examine the origin of the inconsistency, and to fix it. AZ is also grateful for useful discussions and multiple image voting that took place as a community effort in the HFF framework, and to the respective teams that submitted models  -- led by PIs Bradac, Natarajan \& Kneib (CATS), Merten \& Zitrin, Sharon, Williams, Keeton, Bernstein and Diego, and the GLAFIC group. The work uses some scripts from the astronomy \textsc{Matlab} package \citep{OfekMatlab2014ascl.soft07005O}. This work is based on observations obtained with the NASA/ESA Hubble Space Telescope, retrieved from the Mikulski Archive for Space Telescopes (MAST) at the Space Telescope Science Institute (STScI). STScI is operated by the Association of Universities for Research in Astronomy, Inc. under NASA contract NAS 5-26555.



\begin{deluxetable*}{lccccc}
\tablecaption{Multiple Image Systems}
\label{multTable}
\tablecolumns{6}
\tablewidth{0.85\linewidth}
\tablehead{
\colhead{ID
} &
\colhead{R.A
} &
\colhead{DEC.
} &
\colhead{$z_{spec}$ 
} &
\colhead{$z_{phot}$ 
} &
\colhead{Category}
}
\startdata
1.1 & 177.39700 & 22.39600 & 1.488 & & Gold \\
1.2 &177.39942 &22.39743 &1.488&&Gold\\
1.3 &177.40342 &22.40243 &1.488&&Gold\\
1.5 &177.39986 &22.39713 &---&&Silver\\
 \hline
2.1 &177.40242& 22.38975& 1.891&&Gold\\
2.2 &177.40604& 22.39247& 1.891&&Gold\\
2.3& 177.40658& 22.39288& 1.891&&Gold\\
\hline
3.1 &177.39075 &22.39984 &3.129&&Gold\\
3.2 &177.39271 &22.40308 &3.129&&Gold\\
3.3 &177.40129 &22.40718 &3.129&&Gold\\
\hline
4.1 &177.39300 &22.39682 &2.949&&Gold\\
4.2 &177.39438 &22.40073 &2.949&&Gold\\
4.3 &177.40417 &22.40612& 2.949&&Gold\\
\hline
5.1 & 177.39975& 22.39306 &2.80 &  &Gold\\
5.2 & 177.40108& 22.39382 &---& $2.6 \pm 0.1$ &Gold\\
5.3 & 177.40792 &22.40355 &---& $2.8 \pm 0.1$ &Silver\\
\hline
6.1 &177.39971 &22.39254 &---&&Gold\\
6.2 &177.40183 &22.39385 &---& $2.6^{+0.1}_{-2.3}$ &Gold\\
6.3 &177.40804 &22.40250&---&&Silver\\
\hline
7.1 &177.39896 &22.39133&---&$2.6\pm0.1$&Gold\\
7.2 &177.40342 &22.39426&---&$2.7^{+0.1}_{-0.2}$&Gold\\
7.3 &177.40758 & 22.40124&---&$2.3^{+0.1}_{-0.2}$&Gold\\
\hline
13.1 &177.40371 &22.39778 &1.23&&Gold\\
13.2 &177.40283 &22.39665& 1.25&&Gold\\
13.3 &177.40004 &22.39385 &1.23&&Gold\\
\hline
14.1 &177.39167 &22.40348 &3.703 &&Gold\\
14.2 &177.39083 &22.40264 & 3.703&&Gold\\
\hline
110.1 &177.40014 &22.39016 &3.214&&Gold\\
110.2 &177.40402& 22.39289& 3.214&&Gold\\
\enddata
\tablecomments{Multiple images used as constraints in the modeling. These systems comprise the Gold sample, as defined in \citep{Treu2016Refsdal} following a ranking of the images based on votes from various lensing teams. While some differences exist, generally speaking, these systems were used in constructing the models for the comparison work of SN Refsdal \citep{Treu2016Refsdal,Kelly2016reappearance,Rodney2016RefsdalComp}, but some also used the silver sample, not shown in full here. Many of the images in the Table (namely those of systems 1 through 7, and 13) were originally found in our earlier work by previous versions of the LTM model for this cluster \citep[][see also \citep{Smith2009M1149}]{ZitrinBroadhurst2009,Zheng2012Nature,Zitrin2014CLASH25}. Indices, coordinates, and redshifts are taken from  \citet{Finney2018M1149}, following \citet{Treu2016Refsdal}.}
\end{deluxetable*}

\clearpage
\startlongtable
\begin{deluxetable}{lcc}
\tablecaption{Multiply Imaged Knots}
\label{knotTable}
\tablecolumns{3}
\tablewidth{0.99\linewidth}
\tablehead{
\colhead{ID
} &
\colhead{R.A
} &
\colhead{DEC.
}
}
\startdata
\hline
1.1.1 & 177.39702 & 22.39600 \\
1.1.2 & 177.39942 & 22.39743 \\
1.1.3 & 177.40341 & 22.40244 \\
1.1.5* & 177.39986 & 22.39713 \\
\hline
1.2.1 & 177.39661 & 22.39630 \\ 
1.2.2 & 177.39899 & 22.39786 \\ 
1.2.3 & 177.40303 & 22.40268 \\ 
1.2.4 & 177.39777 & 22.39878 \\
1.2.6 & 177.39867 & 22.39824 \\
\hline
1.3.1 & 177.39687 & 22.39621 \\ 
1.3.2 & 177.39917 & 22.39760 \\ 
1.3.3 & 177.40328 & 22.40259 \\
\hline
1.4.1 & 177.39702 & 22.39621 \\ 
1.4.2 & 177.39923 & 22.39748 \\ 
1.4.3 & 177.40339 & 22.40255 \\
\hline
1.5.1 & 177.39726 & 22.39620 \\ 
1.5.2 & 177.39933 & 22.39730 \\ 
1.5.3 & 177.40356 & 22.40252 \\
\hline
1.6.1 & 177.39737 & 22.39616 \\ 
1.6.2 & 177.39945 & 22.39723 \\ 
1.6.3 & 177.40360 & 22.40248 \\
\hline
1.7.1 & 177.39757 & 22.39611 \\ 
1.7.2 & 177.39974 & 22.39693 \\
1.7.3 & 177.40370 & 22.40240 \\
\hline
1.8.1 & 177.39795 & 22.39601 \\ 
1.8.2 & 177.39981 & 22.39675 \\ 
1.8.3 & 177.40380 & 22.40231 \\
\hline
1.9.1 & 177.39803 & 22.39593 \\ 
1.9.2 & 177.39973 & 22.39698 \\ 
1.9.3 & 177.40377 & 22.40225 \\
\hline
1.10.1 & 177.39809 & 22.39585 \\ 
1.10.2 & 177.39997 & 22.39670 \\ 
1.10.3 & 177.40380 & 22.40218 \\
\hline
1.11.2 & 177.40010 & 22.39666 \\ 
1.11.3 & 177.40377 & 22.40204 \\
\hline
1.12.1 & 177.39716 & 22.39521 \\
1.12.2 & 177.40032 & 22.39692 \\
1.12.3 & 177.40360 & 22.40187 \\
\hline
1.13.1 & 177.39697 & 22.39663 \\ 
1.13.2 & 177.39882 & 22.39771 \\ 
1.13.3 & 177.40329 & 22.40282 \\ 
1.13.4 & 177.39791 & 22.39843 \\
\hline
1.14.1 & 177.39712 & 22.39672 \\
1.14.2 & 177.39878 & 22.39763 \\ 
1.14.3 & 177.40338 & 22.40287 \\ 
1.14.4 & 177.39810 & 22.39825 \\
\hline
1.15.1 & 177.39717 & 22.39650 \\ 
1.15.2 & 177.39894 & 22.39751 \\ 
1.15.3 & 177.40344 & 22.40275 \\
\hline
1.16.1 & 177.39745 & 22.39640 \\ 
1.16.2 & 177.39915 & 22.39722 \\
1.16.3 & 177.40360 & 22.40265 \\
\hline
1.17.1 & 177.39815 & 22.39634 \\ 
1.17.2 & 177.39927 & 22.39683 \\ 
1.17.3 & 177.40384 & 22.40256 \\
\hline
1.18.1 & 177.39850 & 22.39610 \\ 
1.18.2 & 177.39947 & 22.39659  \\
1.18.3 & 177.40394 & 22.40240 \\
\hline
1.19.1 & 177.39689 & 22.39576 \\
1.19.2 & 177.39954 & 22.39748 \\
1.19.3 & 177.40337 & 22.40229 \\
1.19.5 & 177.39997 & 22.39710 \\
\hline
1.20.1 & 177.39708 & 22.39572 \\ 
1.20.2 & 177.39963 & 22.39736 \\
1.20.3 & 177.40353 & 22.40223 \\ 
1.20.5 & 177.40000 & 22.39698 \\
\hline
1.21.1 & 177.39694 & 22.39540  \\
1.21.3 & 177.40341 & 22.40200 \\
1.21.5 & 177.40018 & 22.39704 \\
\hline
1.22.1 & 177.39677 & 22.39548 \\ 
1.22.2 & 177.39968 & 22.39749 \\ 
1.22.3 & 177.40328 & 22.40209 \\ 
1.22.5 & 177.40008 & 22.39713 \\
\hline
1.23.1 & 177.39672 & 22.39538 \\ 
1.23.2 & 177.39977 & 22.39749 \\ 
1.23.3 & 177.40324 & 22.40201 \\ 
1.23.5 & 177.40013 & 22.39720 \\
\hline
1.24.1 & 177.39650 & 22.39558 \\ 
1.24.2 & 177.39953 & 22.39775 \\ 
1.24.3 & 177.40301 & 22.40220 \\
\hline
1.25.1 & 177.39657 & 22.39593 \\ 
1.25.3 & 177.40304 & 22.40245 \\
\hline
1.26.1 & 177.39633 & 22.39601 \\ 
1.26.3 & 177.40283 & 22.40260 \\
\hline
1.27.1 & 177.39831 & 22.39628 \\ 
1.27.2 & 177.39933 & 22.39672 \\
\hline
1.28.1 & 177.39860 & 22.39616 \\
1.28.2 & 177.39942 & 22.39655 \\
\hline
1.29.1 & 177.39858 & 22.39586 \\
1.29.2 & 177.39976 & 22.39649 \\
\hline
1.30.1 & 177.39817 & 22.39546 \\ 
1.30.2 & 177.39801 & 22.39523 \\ 
1.30.3 & 177.39730 & 22.39536 \\ 
1.30.4 & 177.39788 & 22.39572 \\ 
\hline
SN1 &177.39823 &22.39563\\
SN2 &177.39772 &22.39578\\
SN3 &177.39737 &22.39553\\
SN4 &177.39781 &22.39518\\
\hline
SX*$^{a}$ &177.40024 &22.39681\\
SY*$^{a}$ &177.40380 &22.40214\\
\enddata
\tablecomments{In addition to the multiple image systems above, we also use individual, multiply imaged knots in the lensed spiral hosting Refsdal ($z=1.49$) as constraints. Also here the IDs and coordinates follow \cite{Finney2018M1149}, based on \cite{Treu2016Refsdal}.\\
$^{*}$ Images with an asterisk are shown for completeness and were not used in the minimization.\\
$^{a}$ Note that for SX and SY there seems to be some discrepancy between the listed coordinates (taken from \cite{Finney2018M1149}) and those we measure in the data ourselves.}
\end{deluxetable}


\end{document}